\pdfoutput=1
\documentclass[twocolumn,prl,superscriptaddress,longbibliography]{revtex4-2}
\usepackage{graphicx,empheq,amssymb,scalerel,tikz}
\renewcommand{\section}[1]{{\par\it #1.---}\ignorespaces}
\usepackage[colorlinks=true, citecolor=blue, urlcolor=blue, linkcolor=blue ]{hyperref}
\usetikzlibrary{svg.path}
\definecolor{orcidlogocol}{HTML}{A6CE39}
\tikzset{
	orcidlogo/.pic={
		\fill[orcidlogocol] svg{M256,128c0,70.7-57.3,128-128,128C57.3,256,0,198.7,0,128C0,57.3,57.3,0,128,0C198.7,0,256,57.3,256,128z};
		\fill[white] svg{M86.3,186.2H70.9V79.1h15.4v48.4V186.2z}
		svg{M108.9,79.1h41.6c39.6,0,57,28.3,57,53.6c0,27.5-21.5,53.6-56.8,53.6h-41.8V79.1z M124.3,172.4h24.5c34.9,0,42.9-26.5,42.9-39.7c0-21.5-13.7-39.7-43.7-39.7h-23.7V172.4z}
		svg{M88.7,56.8c0,5.5-4.5,10.1-10.1,10.1c-5.6,0-10.1-4.6-10.1-10.1c0-5.6,4.5-10.1,10.1-10.1C84.2,46.7,88.7,51.3,88.7,56.8z};}}
\newcommand\orcid[1]{\href{https://orcid.org/#1}{\mbox{\scalerel*{\begin{tikzpicture}[yscale=-1,transform shape]\pic{orcidlogo};\end{tikzpicture}}{|}}}}

\begin{document}
\title{Generalized Quantum Fluctuation Theorem for Energy Exchange}
\author{Wei Wu\orcid{0000-0002-7984-1501}}
\affiliation{Key Laboratory of Quantum Theory and Applications of MoE, Lanzhou Center for Theoretical Physics, and Key Laboratory of Theoretical Physics of Gansu Province, Lanzhou University, Lanzhou 730000, China}
\author{Jun-Hong An\orcid{0000-0002-3475-0729}}
\email{anjhong@lzu.edu.cn}
\affiliation{Key Laboratory of Quantum Theory and Applications of MoE, Lanzhou Center for Theoretical Physics, and Key Laboratory of Theoretical Physics of Gansu Province, Lanzhou University, Lanzhou 730000, China}

\begin{abstract}
The nonequilibrium fluctuation relation is a cornerstone of quantum thermodynamics. It is widely believed that the system-bath heat exchange obeys the famous Jarzynski-W\'{o}jcik fluctuation theorem. However, this theorem is established in the Born-Markovian approximation under the weak-coupling condition. Via studying the energy exchange between a harmonic oscillator and its coupled bath in the non-Markovian dynamics, we establish a generalized quantum fluctuation theorem for energy exchange being valid for arbitrary coupling strength. The Jarzynski-W\'{o}jcik fluctuation theorem is recovered in the weak-coupling limit. We also find the average energy exchange exhibits rich nonequilibrium characteristics when different numbers of system-bath bound states are formed, which suggests a useful way to control the quantum heat. Deepening our understanding of the fluctuation relation in quantum thermodynamics, our result lays the foundation to design high-efficiency quantum heat engines.
\end{abstract}
\maketitle

\section{Introduction}
The rapid development of quantum technologies raises the requirement of creating quantum thermodynamics to describe the emergent physics of quantum machines \cite{Brandao2015,doi:10.1080/00107514.2016.1201896,Goold_2016,DePasquale2018,10.1088/2053-2571/ab21c6,RevModPhys.92.041002}. Concerned with thermodynamic phenomena occurring at the quantum level, quantum thermodynamics aims to rebuild thermodynamics for small systems from the microscopic dynamics of quantum mechanics \cite{Neill2016,PhysRevB.98.134306,PhysRevLett.124.160601,PRXQuantum.2.030202,RODUNER20221} and design revolutionary quantum machines outperforming their classical counterparts by quantum features \cite{Jaramillo_2016,doi:10.1126/science.aad6320,10.1116/5.0083192}. In the exploration of improving the efficacies of quantum processors, finding quantities to measure and define thermodynamic quantities, e.g., heat and work, and understanding the statistics of their fluctuations during the nonequilibrium dynamics are of crucial importance.

Although thermodynamic quantities are deterministic in macroscopic systems, they become stochastic at the microscopic scale owing to the presence of thermal or quantum fluctuations. An important issue is then to reveal their probability distributions and fluctuation relations for a small system out of equilibrium~\cite{RevModPhys.81.1665,RevModPhys.83.771,Seifert_2012,PhysRevLett.78.2690,PhysRevE.60.2721,PhysRevLett.91.110601,PhysRevLett.96.050601,PhysRevLett.97.140603,PhysRevLett.104.090601,PhysRevLett.92.230602,PhysRevLett.99.180601,PhysRevLett.123.110602,PhysRevLett.91.110601,PhysRevLett.108.240603,PhysRevE.73.045101,PhysRevLett.119.100601,PhysRevA.107.052211}. Providing insightful viewpoints to reexamine thermodynamic laws at a nonequilibrium level~\cite{PhysRevLett.89.050601,PhysRevLett.71.2401,PhysRevLett.92.140601}, they may promote the development of quantum heat engines beyond the efficiency bounds set by the equilibrium thermodynamics via considering the fluctuations in microscopic systems~\cite{Martinez2016,Sinitsyn_2011,Holubec_2022,PhysRevE.101.030101,PhysRevE.106.024105}. The Jarzynski equality~\cite{PhysRevLett.78.2690} and Crooks fluctuation theorem \cite{PhysRevE.60.2721} allow one to determine the fluctuation relation of the nonequilibrium work via measuring the equilibrium free energy, which has been experimentally tested~\cite{PhysRevLett.94.180602,Toyabe2010,PhysRevLett.109.180601,An2015,PhysRevLett.120.080602}. Originally proposed by Jarzynski and W\'{o}jcik~\cite{PhysRevLett.92.230602}, the fluctuation theorem for heat bridges a connection between the nonequilibrium heat and equilibrium temperatures. It has been widely used in studying the quantum heat exchange in a relaxation process~\cite{PhysRevE.98.052106,PhysRevE.99.022133,10.1063/1.3655674,e23121602,PhysRevA.100.042119,PhysRevE.108.014138}. However, the Jarzynski-W\'{o}jcik fluctuation theorem is established under the condition that the system-bath coupling is so weak that the Born-Markovian approximation can be safely adopted \cite{PhysRevE.98.052106,PhysRevE.99.022133,10.1063/1.3655674}. It generally breaks down in the non-Markovian dynamics \cite{PhysRevE.108.014138,RevModPhys.88.021002,Rivas_2014,LI20181}. An interesting question naturally arises: Does a generalized fluctuation theorem exist in systems strongly coupled to baths? On the other hand, it has been revealed that, quantum heat machines exhibit higher efficiencies by optimizing the system-bath coupling strength ~\cite{10.1021/acs.jpclett.5b01404,PhysRevLett.124.160601,PhysRevB.98.134306,PhysRevX.10.031015}. This result implies that a controllable nonequilibrium quantum heat is of importance in thermodynamic cycles. This also raises an urgent requirement to develop a mechanism to steer the energy exchange in systems strongly coupled to baths.

\begin{figure}[tbp]
\centering
\includegraphics[width=0.47\textwidth]{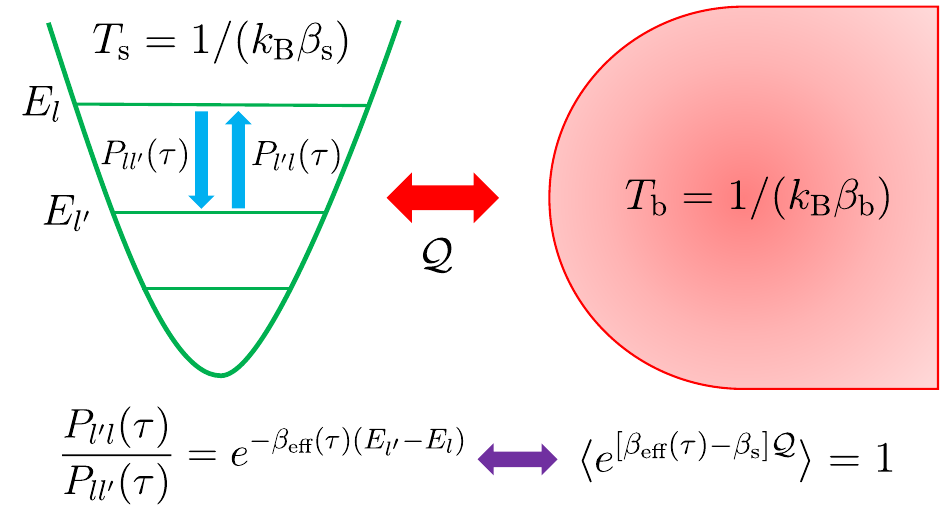}
\caption{Diagrammatic sketch of our result on the energy exchange between a quantum harmonic oscillator and a bath at different initial temperatures in the non-Markovian dynamics. A generalized fluctuation theorem is established by constructing a dynamical detailed balance.}\label{fig:fig1}
\end{figure}

The definition of quantum heat in the strong-coupling regime is controversial \cite{PhysRevE.98.012113,e23121602,Xu_2022,PhysRevLett.127.250601}. For example, Ref. \cite{PhysRevLett.127.250601} defines it via the Hamiltonian of mean force based on an assumption that the total system equilibrates to a Gibbs state at the same temperature as the bath. However, this might be invalidated by the strong coupling ~\cite{PhysRevLett.121.220403,PhysRevResearch.4.023141,PhysRevE.90.022122,PhysRevA.89.012128}. We do not address this ongoing debate here and focus on the statistics of the energy absorbed from or released to the bath \cite{e23121602}. We investigate the energy exchange between a quantum harmonic oscillator and a bath in the non-Markovian dynamics. In the weak-coupling limit, such an energy exchange can be regarded as the quantum heat~\cite{PhysRevE.98.052106,PhysRevE.99.022133}, and its statistics recovers the Jarzynski-W\'{o}jcik fluctuation theorem \cite{PhysRevE.98.052106}. We establish a generalized fluctuation theorem for energy exchange, which is valid for arbitrary coupling strength, by constructing a dynamical detailed balance of the open system. We further find that, in contrast to the previous Born-Markovian results, the average energy exchange sensitively depends on the feature of the energy spectrum of the total system consisting of the system and the bath. With the formation of different numbers of system-bath bound states, the average energy exchange exhibits rich nonequilibrium characteristics. This offers us a method to harvest diverse nonequilibrium quantum heat by engineering the energy-spectrum feature of the total system. 

\section{Jarzynski-W\'{o}jcik fluctuation theorem}
Consider a system interacting with a bath governed by $\hat{H}_{\text{tot}}=\hat{H}_{\text{s}}+\hat{H}_{\text{b}}+\hat{H}_{\text{i}}$, where $\hat{H}_{\text{s}}$, $\hat{H}_{\text{b}}$, and $\hat{H}_{\text{i}}$ are the Hamiltonians of the system, the bath, and their interaction, respectively; see Fig. \ref{fig:fig1}. The system-bath energy exchange can be quantified by two-point measurements of the system energy as $\mathcal{Q}_{l'l}=E_{l'}-E_{l}$, where $E_{l}$ and $E_{l'}$ satisfying $\hat{H}_{\text{s}}=\sum_{l}E_{l}|l\rangle\langle l|$ are the measured energies at $t=0$ and $\tau$, respectively ~\cite{e23121602,PhysRevE.108.014138}. The probability of obtaining $E_{l}$ at $t=0$ is $P_{l}(0)=\langle l|\rho(0)|l\rangle$. The probability of obtaining $E_{l'}$ at $t=\tau$ is $P_{l'l}(\tau)=\text{Tr}[|l'\rangle\langle l'|e^{-i\hat{H}_{\text{tot}}\tau}|l\rangle\langle l|\otimes\rho_{\text{b}}(0)e^{i\hat{H}_{\text{tot}}\tau}]$. Then, the probability distribution of energy exchange reads $\mathcal{P}_\tau(\mathcal{Q})\equiv\sum_{l'l}P_{l'l}(\tau)P_{l}(0)\delta(\mathcal{Q}-\mathcal{Q}_{l'l})$. Its characteristic function is $\chi_{\tau}(\xi)=\int d\mathcal{Q}e^{\xi \mathcal{Q}}\mathcal{P}_\tau(\mathcal{Q})$.

The energy exchange is just the quantum heat in the weak-coupling case ~\cite{PhysRevE.98.052106,PhysRevE.99.022133}. The Jarzynski-W\'{o}jcik fluctuation theorem characterizes the statistics of heat exchange between the system and the bath at different initial temperatures, i.e., $\rho(0)={e^{-\beta_\text{s}\hat{H}_\text{s}}\over\mathcal{Z}(\beta_\text{s})}$ and $\rho_{\text{b}}(0)={e^{-\beta_{\text{b}}\hat{H}_{\text{b}}}\over\text{Tr}(e^{-\beta_{\text{b}}\hat{H}_{\text{b}}})}$, with $\mathcal{Z}(\beta_\text{s})=\text{Tr}(e^{-\beta_\text{s}\hat{H}_\text{s}})$ and $\beta_\text{s/b}$ being their inverse temperatures. It is guaranteed by the detailed-balance condition ${P_{l'l}(\tau)\over P_{ll'}(\tau)}=e^{-\beta_\text{b} (E_{l'}-E_l)}$, under which the characteristic function $\chi_{\tau}(\xi)$ has the permutation symmetry $\chi_{\tau}(\xi)=\chi_{\tau}(\beta_{\text{b}}-\beta_{\text{s}}-\xi)$, and thus the distribution function $\mathcal{P}_{\tau}(\mathcal{Q})$ satisfies $\ln{\mathcal{P}_{\tau}(-\mathcal{Q})\over \mathcal{P}_{\tau}(\mathcal{Q})}=(\beta_{\text{b}}-\beta_{\text{s}})\mathcal{Q}$~\cite{PhysRevLett.92.230602,PhysRevE.98.052106,PhysRevE.99.022133,e23121602}. It readily results in $\langle e^{(\beta_{\text{b}}-\beta_{\text{s}})\mathcal{Q}}\rangle=1$ with $\langle O\rangle=\int d\mathcal{Q}\mathcal{P}_{\tau}(\mathcal{Q})O$~\cite{PhysRevLett.92.230602}. This is the Jarzynski-W\'{o}jcik fluctuation theorem. However, it is valid only in the Born-Markovian approximation under the weak system-bath coupling condition~\cite{PhysRevE.98.052106,PhysRevE.99.022133,10.1063/1.3655674}. With the rapid development of strong-coupling quantum thermodynamics ~\cite{10.1021/acs.jpclett.5b01404,PhysRevLett.124.160601,PhysRevB.98.134306,PhysRevX.10.031015}, this is obviously not sufficient. 

\section{Generalized fluctuation theorem}
Going beyond the weak-coupling condition, we use a prototypical model of open systems to establish a generalized fluctuation theorem for energy exchange in the exact nonequilibrium dynamics. Consider a  harmonic oscillator $\hat{H}_\text{s}=\omega_0\hat{a}^\dag\hat{a}$  coupled to a bath $\hat{H}_\text{b}=\sum_k\omega_k\hat{b}^\dag_k\hat{b}_k$ via $\hat{H}_\text{i}=\sum_k (g_k\hat{a}^\dag\hat{b}_k+\text{h.c.})$. Here, $\hat{a}$ and $\hat{b}_{k}$ are the annihilation operators of the oscillator with frequency $\omega_0$ and the $k$th bath mode with frequency $\omega_{k}$, and $g_{k}$ is their coupling strength. The coupling is characterized by the spectral density $J(\omega)\equiv\sum_{k}|g_{k}|^{2}\delta(\omega-\omega_{k})$ generally taking an Ohmic-family form $J(\omega)=\eta\omega^{s}\omega_{c}^{1-s}e^{-\omega/\omega_{c}}$, where $\eta$ is a dimensionless coupling constant, $\omega_{c}$ is a cutoff frequency, and $s$ is an Ohmicity index. Provided that the bath is initially in the Gibbs state $\rho_\text{b}(0)$, an exact master equation of the oscillator is derived by the Feynman-Vernon's influence functional method as~\cite{PhysRevA.76.042127,PhysRevE.90.022122,PhysRevApplied.17.034073}
\begin{eqnarray}
\dot{\rho}(t)=i\Omega_t[\rho(t),\hat{a}^{\dagger}\hat{a}]+\big\{{\gamma_t^{\beta}}\check{\mathcal L}_{\hat{a}^\dag}+\big{[}{\gamma_t^{\beta}}+\gamma_t\big{]}\check{\mathcal L}_{\hat{a}}\big\}\rho(t),\label{eq:eq4}
\end{eqnarray}
where $\Omega_t=-\text{Im}[\dot{u}(t)/u(t)]$ is the renormalized frequency, $\gamma_t=-\text{Re}[\dot{u}(t)/u(t)]$ and $\gamma_t^{\beta}=\dot{v}(t)/2+v(t)\gamma_t$ are the bath-induced dissipation and noise coefficients, and $\check{\mathcal L}_{\hat{o}}\rho=2\hat{o}\rho\hat{o}^\dag-\{\hat{o}^\dag\hat{o},\rho\}$ is the Lindblad superoperator. The functions $u(t)$ and $v(t)$ satisfy
\begin{eqnarray}
\dot{u}(t)+i\omega_{0}u(t)+\int_{0}^{t}d t_1\mu(t-t_1)u(t_1)=0,\label{eq:eq5}\\
v(t)=\int_{0}^{t}dt_{1}\int_{0}^{t}dt_{2}u^{*}(t_{1})\nu(t_{1}-t_{2})u(t_{2}),\label{eq:eq6}
\end{eqnarray}
where $u(0)=1$, $\mu(t)=\int_{0}^{\infty}d\omega J(\omega)e^{-i\omega t}$, and $\nu(t)=\int_{0}^{\infty}d\omega {J(\omega)e^{-i\omega t}\over e^{\beta_{\text{b}}\omega}-1}$. The convolution in Eqs. \eqref{eq:eq5} and \eqref{eq:eq6} renders the dynamics non-Markovian, with all the memory effects self-consistently incorporated into the time-dependent coefficients in Eq. \eqref{eq:eq4}. It can be proven that the characteristic function equals \cite{SupplementalMaterial}
\begin{eqnarray}
\chi_{\tau}(\xi)=\frac{\mathcal{Z}(\beta_{\text{s}}+\xi)}{\mathcal{Z}(\beta_{\text{s}})}[1+(1-e^{\xi\omega_{0}})\bar{n}(\beta _\text{s}+\xi,\tau)]^{-1},\label{ddfc}
\end{eqnarray}
where $\bar{n}(\beta _\text{s}+\xi,\tau)={\left\vert u(\tau)\right\vert ^{2}\over e^{\omega_0(\beta_\text{s}+\xi)}-1}+v(\tau)$. Equation \eqref{ddfc} is exact and fully characterizes the statistics of energy exchange beyond the Born-Markovian approximation. The average energy exchange is calculated via $\langle\mathcal{Q}(\tau)\rangle=\partial_{\xi}\chi_{\tau}(\xi)|_{\xi=0}$. An obvious conclusion of Eq. \eqref{ddfc} is $\chi_{\tau}(\xi)\neq\chi_{\tau}(\beta_\text{b}-\beta_{\text{s}}-\xi)$. It implies that our system in the non-Markovian dynamics violates the conventional detailed balance and reveals that the Jarzynski-W\'{o}jcik fluctuation theorem is not valid anymore.

An effective inverse temperature $\beta_\text{eff}(\tau)$ to dynamically establish a generalized detailed balance can be constructed. Solving Eq. \eqref{eq:eq4} under $\rho(0)=|l\rangle\langle l|$, we obtain the transition probability $P_{l'l}(\tau)$ from $|l\rangle$ to $|l'\rangle$ and the transition probability $P_{ll'}(\tau)$ from $|l'\rangle$ to $|l\rangle$ \cite{SupplementalMaterial}. It is remarkable to find that they satisfy a generalized detailed balance ${P_{l'l}(\tau)\over P_{ll'}(\tau)}= e^{-\beta_\text{eff}(\tau)(E_{l'}-E_l)}$
if we define an effective inverse temperature $\beta_\text{eff}(\tau)\equiv\omega_0^{-1}\ln[1+{1-|u(\tau)|^2\over v(\tau)}]$. Being valid during the whole dynamics for arbitrary coupling strength, this generalized detailed balance endows Eq. \eqref{ddfc} with a symmetry $\chi_\tau(\xi)=\chi_\tau[\beta_\text{eff}(\tau)-\beta_\text{s}-\xi]$. It results in the fluctuation theorem $\ln\frac{\mathcal{P}_{\tau}(-\mathcal{Q})}{\mathcal{P}_{\tau}(\mathcal{Q})}=[\beta_{\text{eff}}(\tau)-\beta_{\text{s}}]\mathcal{Q}$, which connects the microscopic distribution function and the measurable energies. It leads to \cite{SupplementalMaterial}
\begin{equation}
    \langle e^{[\beta_{\text{eff}}(\tau)-\beta_\text{s}]\mathcal{Q}}\rangle=1.\label{gftnm}
\end{equation}Working in the full-parameter space, Eq. \eqref{gftnm} is a generalization of the Jarzynski-W\'{o}jcik fluctuation theorem to the non-Markovian dynamics. It bridges a connection between the nonequilibrium energy exchange and the dynamical temperature, with the non-Markovian effect automatically incorporated. It renews our understanding of the fluctuation relation beyond the conventional Born-Markovian and detail-balanced dynamics.

In the special case of the weak coupling, we recover the Jarzynski-W\'{o}jcik fluctuation theorem for quantum heat. The Born-Markovian approximate solutions are $u_\text{MA}(\tau)= e^{-\kappa \tau-i[\omega_{0}+\Delta(\omega_{0})]\tau}$ and $v_\text{MA}(\tau)=\bar{n}(\beta_{\text{b}},0)(1-e^{-2\kappa\tau})$ with $\kappa=\pi J(\omega_{0})$, $\Delta(\omega_{0})=\text{P}\int_{0}^{\infty}\frac{J(\omega)}{\omega_{0}-\omega}d\omega$, and $\text{P}$ being the Cauchy principal value~\cite{PhysRevE.90.022122}. We then have
\begin{eqnarray}
\beta^\text{MA}_\text{eff}(\tau)&=&\beta_\text{b},\label{makeff}\\
    \chi^\text{MA}_\tau(\xi)&=&{e^{\xi\omega_0}(e^{\beta_\text{s}\omega_0}-1)\over e^{(\beta_\text{s}+\xi)\omega_0}-1+V(\xi,\tau)(e^{\xi\omega_0}-1) },\label{marchii}
\end{eqnarray}
where $V(\xi,\tau)=v_\text{MA}(\tau)[1-e^{(\beta_\text{s}+\xi)\omega_0}]-|u_\text{MA}(\tau)|^2$. Equation \eqref{makeff} reduces Eq. \eqref{gftnm} to the Jarzynski-W\'{o}jcik fluctuation theorem \cite{PhysRevLett.92.230602}. Equation \eqref{marchii} is the same as the one in Ref. \cite{PhysRevE.98.052106}. We obtain the average energy exchange as \begin{eqnarray}
    \langle\mathcal{Q}_{\text{MA}}(\tau)\rangle=\omega_0\Big[v_\text{MA}(\tau)+{1-|u_\text{MA}(\tau)|^2\over 1-e^{\beta_\text{s}\omega_0}}\Big].\label{makht}
\end{eqnarray}
Its long-time limit is $\langle\mathcal{Q}_{\text{MA}}(\infty)\rangle=\omega_{0}[\bar{n}(\beta_{\text{b}},0)-\bar{n}(\beta_{\text{s}},0)]$, which depends only on the temperatures of the system and the bath, but not on the details of $J(\omega)$. Matching with the average quantum heat in Refs.~\cite{PhysRevE.98.052106,PhysRevE.99.022133,10.1063/1.3655674,e23121602,Xuereb_2015}, the result is universal for any heat-exchange process experiencing a canonical thermalization~\cite{e23121602}.

\begin{figure}
\centering
\includegraphics[width=0.47\textwidth]{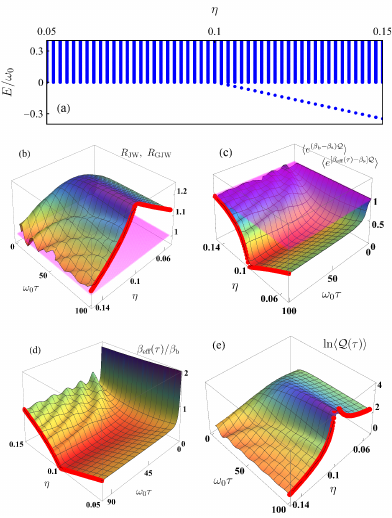}
\caption{(a) Energy spectrum. (b) Evolution of $R_{\text{JW}}(\xi)\equiv\chi_{\tau}(\beta_{\text{b}}-\beta_{\text{s}}-\xi)/\chi_\tau(\xi)$ (the rainbow surface) and $R_{\text{GJW}}(\xi)\equiv\chi_{\tau}(\beta_{\text{eff}}(\tau)-\beta_{\text{s}}-\xi)/\chi_\tau(\xi)$ (the magenta surface) in different $\eta$ when $\xi=2.25$. Evolution of $\langle e^{(\beta_{\text{b}}-\beta_{\text{s}})\mathcal{Q}}\rangle$ [the rainbow surface in (c)], $\langle e^{(\beta_{\text{eff}}(\tau)-\beta_{\text{s}})\mathcal{Q}}\rangle$ [the magenta surface in (c)], (d) $\beta_\text{eff}(\tau)$, and (e) average energy exchange $\langle\mathcal{Q}(\tau)\rangle$ in different $\eta$. The red lines in (b) and (c) are from Eq. \eqref{extchiinf}. The red lines in (d) and (e) are from Eqs. \eqref{exteftem} and (\ref{extht}). We use $s=1$, $\omega_{c}=10\omega_{0}$, $\beta_{\text{s}}=1.2\omega_{0}^{-1}$, and $\beta_{\text{b}}=0.2\omega_{0}^{-1}$.}\label{fig:fig2}
\end{figure}
In the non-Markovian case, Eqs. \eqref{ddfc} and \eqref{gftnm} are obtainable via numerically solving Eqs. \eqref{eq:eq5} and \eqref{eq:eq6}. However, we may derive their steady-state forms. $u(t)$ is the inverse Laplace transform of $\tilde{u}(z)=[z+i\omega_{0}+\int_0^\infty d\omega{J(\omega)\over z+i\omega}]^{-1}$, which requires that we find the poles of $\tilde{u}(z)$ from
\begin{equation}\label{eq:eq9}
y(E)\equiv\omega_{0}-\int_0^\infty d\omega{J(\omega)\over\omega-E} =E, (E=iz).
\end{equation}
First, the roots of Eq. \eqref{eq:eq9} are the eigenenergies of $\hat{H}_\text{tot}$ in the single-excitation subspace. To prove this, we expand the eigenstate as $|\Phi\rangle=(x\hat{a}^{\dagger}+\sum_{k}y_{k}\hat{b}_{k}^{\dagger})|0,\{0_k\}\rangle$. The substitution of it into $\hat{H}_\text{tot}|\Phi\rangle=E|\Phi\rangle$ leads to Eq.~(\ref{eq:eq9}). Thus, $u(t)$ is governed by the energy-spectrum characteristic of $\hat{H}_\text{tot}$. Second, Eq.~(\ref{eq:eq9}) has an isolated root $E_\text{b}<0$ provided $y(0)<0$ because $y(E)$ is a decreasing function when $E<0$. We call the eigenstate of $E_\text{b}$ the bound state. Since $y(E)$ is not analytic when $E>0$ due to the divergence of the integral in $y(E)$, Eq.~(\ref{eq:eq9}) has infinite roots forming an energy band in this regime. The condition of forming the bound state is evaluated via $y(0)<0$ as $\omega_{0}<\eta\omega_c\Gamma(s)$, with $\Gamma(s)$ being the Euler's gamma function. Using the residue theorem, we obtain $u(t)=Ze^{-iE_\text{b} t}+\int_{0}^{\infty}\Theta(\omega)e^{-i\omega t}d\omega$, where $Z\equiv[1+\int_0^\infty{J(\omega)d\omega\over(E_\text{b}-\omega)^2}]^{-1}$ and $\Theta(\omega)=\frac{J(\omega)}{[\omega-\omega_{0}-\Delta(\omega)]^{2}+[\pi J(\omega)]^2}$~\cite{PhysRevA.103.L010601}. The second term tends to zero with time due to the out-of-phase interference. Thus, when the bound state is absent, $u(\infty)=0$ means a complete decoherence; when the bound state is present, $u(\infty)= Ze^{-iE_\text{b}t}$ implies a decoherence suppression. We can derive $v(\infty)=\int_0^\infty{J(\omega)\over e^{\beta_\text{b}\omega}-1}\big[\Theta(\omega)+{Z^2\over (\omega-E_\text{b})^2}\big]d\omega$ and
\begin{eqnarray}
 \beta_\text{eff}(\infty)&=&\omega_0^{-1}\ln\Big[1+{1-Z^2\over v(\infty)}\Big],\label{exteftem}\\
 \chi_\infty(\xi)&=&{e^{\xi\omega_0}(e^{\beta_\text{s}\omega_0}-1)\over e^{(\beta_\text{s}+\xi)\omega_0}-1+\mathcal{V}(\xi)(e^{\xi\omega_0}-1)},\label{extchiinf}
\end{eqnarray}
where $\mathcal{V}(\xi)=v(\infty)[1-e^{(\beta_\text{s}+\xi)\omega_0}]-Z^2$. The average energy exchange  is calculated from Eq. \eqref{extchiinf} as
\begin{equation}
 \langle\mathcal{Q}(\infty)\rangle=\omega_0\Big[v(\infty)+{1-Z^2\over 1-e^{\beta_\text{s}\omega_0}}\Big]. \label{extht}
\end{equation}
It is interesting to see that, in contrast to Eqs. \eqref{makeff}-\eqref{makht}, Eqs. \eqref{exteftem}-\eqref{extht} exhibit a sensitive dependence on $J(\omega)$ and the energy spectrum of $\hat{H}_\text{tot}$. Accompanying a sudden appearance of the nonzero $Z$ at the critical point of forming the bound state, all of them have an abrupt change. Thus, the energy exchange and its distribution exhibit diverse nonequilibrium behaviors depending on whether the bound state is present or not. This result stems from the noncanonical thermalization in the strong-coupling regime ~\cite{PhysRevLett.121.220403,PhysRevResearch.4.023141,PhysRevE.90.022122,PhysRevA.89.012128} and manifests the breakdown of the Born-Markovian approximate heat statistics.

We verify our result by taking the Ohmic spectral density as an example. The energy spectrum in Fig. \ref{fig:fig2}(a) by numerically solving Eq. \eqref{eq:eq9} confirms our expectation that the bound state is present when $\eta>\omega_0/\omega_c$. The ratios $R_\text{JW}(\xi)\equiv\chi_\tau(\beta_\text{b}-\beta_\text{s}-\xi)/\chi_\tau(\xi)$ and $R_\text{GJW}(\xi)\equiv\chi_\tau[\beta_\text{eff}(\tau)-\beta_\text{s}-\xi]/\chi_\tau(\xi)$ for an arbitrarily chosen $\xi$ in Fig. \ref{fig:fig2}(b) show that $R_\text{JW}(\xi)$ exhibits a dramatic deviation from 1, while $R_\text{GJW}(\xi)$ remains 1 during the whole evolution. It verifies that $\beta_\text{eff}(\tau)$ recovers the permutation symmetry of $\chi_\tau(\xi)$ in the non-Markovian dynamics. The steady-state results match well with Eq. \eqref{extchiinf}. Figure \ref{fig:fig2}(c) indicates that $\langle e^{(\beta_{\text{b}}-\beta_{\text{s}})\mathcal{Q}}\rangle$ exhibits an obvious deviation from 1, while $\langle e^{[\beta_{\text{eff}}(\tau)-\beta_{\text{s}}]\mathcal{Q}}\rangle$ remains 1. $\beta_\text{eff}(\tau)$ guaranteeing the generalized detailed balance in Fig. \ref{fig:fig2}(d) shows not only a dramatic difference from the Born-Markovian result in Eq. \eqref{makeff} but also a peak at the critical point of forming the bound state, which has been used to realize a quantum thermometry \cite{PhysRevApplied.17.034073}. All the results prove the breakdown of the conventional Jarzynski-W\'{o}jcik fluctuation theorem and the validity of our generalized one. The average energy exchange $\langle\mathcal{Q}(\tau)\rangle$ in Fig.~\ref{fig:fig2}(e) exhibits a strong dependence on the formation of the bound state, which is in sharp contrast to the constant result in Eq. \eqref{makht}. This offers us with an insightful guideline to harvest the maximal energy via engineering the energy-spectrum characteristic.  

\begin{figure}
\centering
\includegraphics[width=0.47\textwidth]{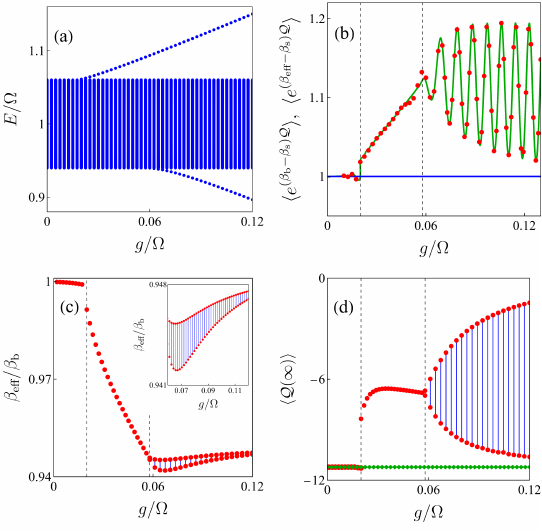}
\caption{(a) Energy spectrum of a resonator coupled to a resonator array with $N=500$. (b) Steady-state $\langle e^{(\beta_{\text{b}}-\beta_{\text{s}})\mathcal{Q}}\rangle$ (the red circles) and $\langle e^{[\beta_{\text{eff}}(\tau)-\beta_{\text{s}}]\mathcal{Q}}\rangle$ (the blue solid line) in different $g$ at $\tau=400\Omega^{-1}$. The green solid lines are from $u_{\text{TBS}}(\infty)$ and $v_{\text{TBS}}(\infty)$. Steady-state $\beta_\text{eff}(\tau)$ (c) and $\langle\mathcal{Q}(\tau)\rangle$ (d), with the maxima and minima of their lossless oscillations marked by the red dots, in different $g$ near $\tau=600\Omega^{-1}$. The green rhombuses in (d) are the Born-Markovian result in the long-time limit of Eq. (\ref{makht}). The black dashed lines mark the positions of forming different numbers of the bound states. We use $\beta_{\text{s}}=\Omega^{-1}$, $\beta_{\text{b}}=5\Omega^{-1}$, $\omega_{0}=1.05\Omega$, and $\zeta=0.03\Omega$.}\label{fig:fig3}
\end{figure}
\section{Physical realization}
A promising platform to test our prediction is a resonator as the system interacting with an array of coupled resonators as the bath. It can be realized in an optical resonator system~\cite{Hafezi2013}, a microdisk cavity system~\cite{10.1063/1.2356892}, a photonic crystal system~\cite{PhysRevB.86.195312,Hennessy2007,Notomi2008}, and optical waveguides~\cite{PhysRevLett.110.076403,Rechtsman2013}. The bath and the interaction Hamiltonians are $\hat{H}_{\text{b}}=\sum_{j=1}^{N}[\Omega\hat{b}_{j}^{\dagger}\hat{b}_{j}+\zeta(\hat{b}_{j}^{\dagger}\hat{b}_{j+1}+\hat{b}_{j+1}^{\dagger}\hat{b}_{j})]$ and
$\hat{H}_{\text{i}}=g(\hat{a}^{\dagger}\hat{b}_{1}+\hat{a}\hat{b}_{1}^{\dagger})$. The dispersion relation reads $\omega_{k}=\Omega+2\zeta\cos k$, which shows a finite bandwidth $4\zeta$ centered at $\Omega$. The spectral density is $J(\omega)={g^{2}\over2\pi\zeta^{2}}\sqrt{4\zeta^{2}-(\omega-\Omega)^{2}}$.
The band-gap structure of the bath endows the platform with a strong non-Markovian effect. It is interesting to see from Fig.~\ref{fig:fig3}(a) that two bound states $E_{b}^{\pm}$ at most could be formed. According to our bound-state analysis, we have
$u_{\text{TBS}}(\infty)=\sum_{n=\pm}Z_{n}e^{-iE_\text{b}^{n}t}$ with $Z_{n}=[1+\int {J(\omega)d\omega \over(E_\text{b}^{n}-\omega)^{2}}]^{-1}$ and $v_{\text{TBS}}(\infty)=\int{J(\omega)\over e^{\beta_\text{b}\omega}-1}\big[\Theta(\omega)+\sum_{n,m=\pm}{Z_{m}Z_{n}\cos[(E_\text{b}^{m}-E_\text{b}^{n})t]\over (\omega-E_\text{b}^{m})(\omega-E_\text{b}^{n})}\big]d\omega$ in the presence of two bound states. Figure~\ref{fig:fig3}(b) confirms that our generalized fluctuation theorem is strictly satisfied, while the original one breaks down. Moreover, it is obviously seen from Figs. ~\ref{fig:fig3}(c) and ~\ref{fig:fig3}(d) that the steady-state $\beta_\text{eff}(\infty)$ and $\langle\mathcal{Q}(\infty)\rangle$ exhibit a sensitive dependence on the formation of the bound states, while the Born-Markovian approximate result is relevant to neither the spectral density nor the energy spectrum. Their real-time dynamics are discussed in \cite{SupplementalMaterial}. It is remarkable to see that, with the presence of two bound states, both $\beta_\text{eff}(\tau)$ and $\langle\mathcal{Q}(\tau)\rangle$ exhibit a lossless oscillation. This endows them with rich nonequilibrium characteristics, which can be widely tunable by using the well-developed quantum reservoir-engineering technique \cite{PhysRevLett.77.4728,doi:10.1126/science.1261033,PhysRevLett.116.240503,Brown2022}. These characteristics may be beneficial for realizing a quantum heat engine.

\section{Discussion and conclusion}
It is noted that, although only the harmonic oscillator system is studied, our method to establish the generalized fluctuation theorem via constructing the dynamical detailed balance is hopefully applicable to the quantum Brownian motion \cite{e23121602,PhysRevE.108.014138} and the dissipative two-level system~\cite{10.1063/5.0095549}. The rich nonequilibrium characteristics of the average energy exchange endowed by the formation of different numbers of system-bath bound states are testable in state-of-the-art quantum-optics experiments because the bound state and its dynamical effect have been observed in circuit quantum electrodynamics architectures~\cite{Liu2017} and matter-wave systems~\cite{Krinner2018,Kwon2022}. These experimental progress provides a strong support to verify our generalized quantum fluctuation theorem.

In summary, we have investigated the statistics of energy exchange of the open system beyond the widely used Born-Markovian approximation. It has been revealed that the original Jarzynski-W\'{o}jcik fluctuation theorem breaks down in the non-Markovian dynamics. A generalized fluctuation theorem for energy exchange is established in the non-Markovian dynamics by recovering the dynamical detailed balance, which is expected to supply us a useful method to uncover the statistics of the energy exchange and even the quantum heat in arbitrary open systems. We have also found that the average energy exchange depends sensitively on the spectral density and the energy spectrum of the total system-bath system. This offers us a highly tunable method to control energy exchange by using the reservoir-engineering technique. Expanding our understanding of the nonequilibrium thermodynamics of open quantum systems, our results lay a foundation for optimizing the performance of quantum heat engines~\cite{PhysRevE.105.044141,Wiedmann_2020,PhysRevResearch.4.033233,PhysRevResearch.5.023066}.

\section{Acknowledgments}
This work is supported by the National Natural Science Foundation of China (Grants No. 12375015, No. 12275109, and No. 12247101) and the Innovation Program for Quantum Science and Technology (Grant No. 2023ZD0300904) of China.
	
\bibliography{reference}
\end{document}


\title{Supplemental Material for ``Generalized Quantum Fluctuation Theorem for Energy Exchange''}
\author{Wei Wu\orcid{0000-0002-7984-1501}}
\affiliation{Key Laboratory of Quantum Theory and Applications of MoE, Lanzhou Center for Theoretical Physics, and Key Laboratory of Theoretical Physics of Gansu Province, Lanzhou University, Lanzhou 730000, China}
\author{Jun-Hong An\orcid{0000-0002-3475-0729}}
\email{anjhong@lzu.edu.cn}
\affiliation{Key Laboratory of Quantum Theory and Applications of MoE, Lanzhou Center for Theoretical Physics, and Key Laboratory of Theoretical Physics of Gansu Province, Lanzhou University, Lanzhou 730000, China}
\maketitle

\section{Characteristic function}
The system-bath energy exchange is defined via the two-time measurement to energy of the system~\cite{e23121602}. Consider that the initial state of the open system is a Gibbs state
$\rho(0)={e^{-\beta_{\text{s}}\omega_{0}\hat{a}^{\dagger}\hat{a}}}/{\mathcal{Z}(\beta_{\text{s}})}$, with $\mathcal{Z}(\beta_\text{s})=\text{Tr}e^{-\beta_{\text{s}}\omega_{0}\hat{a}^{\dagger}\hat{a}}$.
A measurement on its energy at the initial time obtains  $E_{l}=l\omega_{0}$ in a probability
$P_{l}(0)=e^{-\beta_{\text{s}}\omega_{0}l}/\mathcal{Z}(\beta_{\text{s}})$ and collapses the state of the total system into $\rho_\text{tot}(0)=|l\rangle\langle l|\otimes\rho_{\text{b}}(0)$. After the time evolution governed by $\hat{H}_\text{tot}=\hat{H}_\text{s}+\hat{H}_\text{b}+\hat{H}_\text{int}$ within a time duration $\tau$, a second measurement to the energy of the open system is made. The probability of obtaining the eigenenergy $E_{l'}=l'\omega_{0}$ under the condition that the first measurement result is $E_l$ is
\begin{equation}\label{eq:eqs3}
P_{l'l}(\tau)=\langle l'|\text{Tr}_{\text{b}}[\hat{U}(\tau)|l\rangle\langle l|\otimes\rho_{\text{b}}(0)\hat{U}^{\dagger}(\tau)]|l'\rangle,
\end{equation}where $\hat{U}(\tau)=e^{-i\hat{H}_{\text{tot}}\tau}$.
The energy exchange during this process is defined as $\mathcal{Q}_{l'l}=(l'-l)\omega_0$, whose probability distribution reads $\mathcal{P}_\tau(\mathcal{Q})=\sum_{l,l'}P_l(0)P_{l'l}(\tau)\delta(\mathcal{Q}-\mathcal{Q}_{l'l})$.
Its corresponding characteristic function is defined as $\chi_\tau(\xi)=\int d\mathcal{Q} e^{\xi\mathcal{Q}}\mathcal{P}_\tau(\mathcal{Q})$. Using Eq. \eqref{eq:eqs3} and $P_{l}(0)=e^{-\beta_{\text{s}}\omega_{0}l}/\mathcal{Z}(\beta_{\text{s}})$, we can prove that $\chi_\tau(\xi)$ is rewritten as
\begin{eqnarray}\label{eq:eqs4}
\chi_{\tau}(\xi)&=&\sum_{l',l}P_{l'l}(\tau)P_{l}(0)e^{\xi\omega_{0}(l'-l)}\nonumber\\
&=&{\mathcal{Z}(\beta_\text{s}+\xi)\over \mathcal{Z}(\beta_{\text{s}})}\sum_{l'}e^{\xi\omega_{0}l'}\langle l'|\rho(\tau)|l'\rangle,\label{smchat}
\end{eqnarray}where $\rho(\tau)=\text{Tr}_{\text{b}}[\hat{U}(\tau)\tilde{\rho}(0)\otimes\rho_{\text{b}}(0)\hat{U}^{\dagger}(\tau)]$ is the evolved state of the open system under the initial condition $\tilde{\rho}(0)=e^{-(\xi+\beta_\text{s})\omega_0\hat{a}^\dag\hat{a}}/\mathcal{Z}(\xi+\beta_\text{s})$.

The exact form of $\rho(t)$ can be analytically evaluated by the Feynman-Vernon's path-integral influence function method.  The evolution of
the state in the coherent-state representation \cite{PhysRevE.90.022122} reads
\begin{eqnarray}
\rho (\bar{\alpha}_{f},\alpha _{f}^{\prime };t) &=&\int d\mu (\alpha_{i})d\mu (\alpha _{i}^{\prime })\mathcal{J}(\bar{\alpha}_{f},\alpha_{f}^{\prime };t|\bar{\alpha}_{i}^{\prime },\alpha _{i};0)  \nonumber \\
&&\times \rho (\bar{\alpha}_{i},\alpha _{i}^{\prime };0),  \label{timeevo}
\end{eqnarray}
where $\rho (\bar{\alpha}_{f},\alpha _{f}^{\prime };t)\equiv \langle \bar{\alpha}_{f}|\rho \left( t\right) |\alpha _{f}^{\prime }\rangle $ and
\begin{equation}
\rho (\bar{\alpha}_{i},\alpha _{i}^{\prime };0)=\langle \bar{\alpha}_{i}|\tilde{\rho}(0)|\alpha _{i}^{\prime }\rangle={\exp[e^{-(\xi+\beta_\text{s})\omega_0}\bar{\alpha}_i\alpha_l']\over \mathcal{Z}(\beta_\text{s}+\xi)},\label{smint} \end{equation} with $|\alpha \rangle =e^{\alpha \hat{a}^{\dag }}|0\rangle $ and $d\mu (\alpha )=e^{-|\alpha |^{2}}d^{2}\alpha /\pi $. After eliminating the degrees of freedom of the bath, the propagation
functional $\mathcal{J}(\bar{\alpha}_{f},\alpha _{f}^{\prime };t|\bar{\alpha}_{i}^{\prime },\alpha _{i};0)$ reads
\begin{eqnarray}
&&\mathcal{J}(\bar{\alpha}_{f},\alpha _{f}^{\prime };t|\bar{\alpha}_{i}^{\prime },\alpha _{i};0)=M(t)\exp \{J_{1}(t)\bar{\alpha}_{f}\alpha _{i}\nonumber \\
&&~~~+J_{2}\left( t\right) \bar{\alpha}_{f}\alpha _{f}^{\prime}-[J_{3}\left( t\right) -1]\bar{\alpha}_{i}^{\prime }\alpha _{i}+J_{1}^{\ast}(t)\bar{\alpha}_{i}^{\prime }\alpha _{f}^{\prime }\},~~~~  \label{prord}
\end{eqnarray}%
where
\begin{eqnarray}
M(t) &=&[1+v(t)]^{-1},~J_{1}(t)=M(t)u(t),\label{smmm} \\
J_{2}(t) &=&M(t)v(t),~J_{3}(t)=M(t)|u(t)|^{2}.\label{smjj}
\end{eqnarray}%
The functions $u(t)$ and $v(t)$ satisfy
\begin{eqnarray}
\dot{u}(t)+i\omega_{0}u(t)+\int_{0}^{t}d\tau\mu(t-\tau)u(\tau)=0,\label{eq:eq5}\\
v(t)=\int_{0}^{t}dt_{1}\int_{0}^{t}dt_{2}u^{*}(t_{1})\nu(t_{1}-t_{2})u(t_{2}),\label{eq:eq6}
\end{eqnarray}
where $u(0)=1$, $\mu(t)=\int_{0}^{\infty}d\omega J(\omega)e^{-i\omega t}$, and $\nu(t)=\int_{0}^{\infty}d\omega {J(\omega)\over e^{\beta_\text{b}\omega}-1}e^{-i\omega t}$. Substituting Eq. \eqref{prord} into Eq. \eqref{timeevo} and performing the Gaussian integration under the initial condition \eqref{smint}, we obtain
\begin{eqnarray}
\rho (\bar{\alpha}_{f},\alpha _{f}^{\prime };t) &=&\frac{1}{1+\bar{n}(\beta _\text{s}+\xi,t)}\exp [\frac{\bar{n}(\beta _\text{s}+\xi,t)\bar{\alpha}_{f}\alpha _{f}^{\prime }}{\bar{n}(\beta _\text{s}+\xi,t)+1}],\nonumber\\
\end{eqnarray}
where $\bar{n}(\beta _\text{s}+\xi,t)={\left\vert u(t)\right\vert ^{2}\over e^{\omega_0(\beta_\text{s}+\xi)}-1}+v(t)$.
Remembering that the density matrix is expressed as $\rho (t)=\int d\mu
(\alpha _{f})d\mu (\alpha _{f}^{\prime })\rho (\bar{\alpha}_{f},\alpha
_{f}^{\prime };t)|\alpha _{f}\rangle \langle \bar{\alpha}_{f}^{\prime }|$,
we obtain%
\begin{equation}
\rho (t)=\sum_{m=0}^{\infty }\frac{[\bar{n}(\beta _\text{s}+\xi,t)]^{m}}{[1+\bar{n}(\beta _\text{s}+\xi,t)]^{1+m}}|m\rangle \langle m|.\label{smdst}
\end{equation}%
Substituting Eq. \eqref{smdst} into Eq. \eqref{smchat}, we obtain
\begin{equation}\label{eq:eqs11}
\begin{split}
\chi_{\tau}(\xi)=&\frac{\mathcal{Z}(\beta_{\text{s}}+\xi)}{\mathcal{Z}(\beta_{\text{s}})}[1+(1-e^{\xi\omega_{0}})\bar{n}(\beta _\text{s}+\xi,\tau)]^{-1}.
\end{split}
\end{equation}

\section{Generalized quantum fluctuation theorem for energy exchange}
A generalized detailed balance for the non-Markovian dynamics can be established as follows. First, we calculate the evolved state of the open system under the condition $\rho(0)=|l\rangle\langle l|$. After expanding $\rho(0)$ in the coherent-state representation as $\rho(\bar{\alpha}_i,\alpha_i';0)=(\bar{\alpha}_i\alpha_i')^l/l!$, substituting it into Eq. \eqref{timeevo}, and performing the Gaussian integration, we obtain
\begin{eqnarray}
    \rho(\bar{\alpha}_f,\alpha_f';t)&=&M(t)[1-J_3(t)]^le^{J_{2}\left( t\right) \bar{\alpha}_{f}\alpha
_{f}^{\prime }}\sum_{m=0}^{l}\frac{C_{l}^{m}}{m!}\nonumber\\
&&\times[\frac{|J_{1}(t)|^{2}}{%
1-J_{3}(t)}]^{m}(\bar{\alpha}_{f}\alpha _{f}^{\prime })^{m}.
\end{eqnarray}Then the transition probability from the initial state $|l\rangle$ to $|l'\rangle$ at time $\tau$ can be calculated as
\begin{eqnarray}
    P_{l'l}(\tau)&=&\int d\mu
(\alpha _{f})d\mu (\alpha _{f}^{\prime })\rho (\bar{\alpha}_{f},\alpha
_{f}^{\prime };\tau)\langle l'|\alpha _{f}\rangle \langle \bar{\alpha}_{f}^{\prime }|l'\rangle\nonumber\\
&=&M(t)J_{2}(\tau)^{l^{\prime
}}[1-J_{3}(\tau)]^{l}\sum_{m=0}^{\min(l,l')}C_{l}^{m}C_{l^{\prime }}^{m}\nonumber\\
&&\times[\frac{|J_{1}(\tau)|^{2}}{[1-J_{3}(\tau)]J_{2}(\tau)}]^{m}.
\end{eqnarray}
A similar form of the transition probability $P_{ll'}(\tau)$ from the initial state $|l'\rangle$ to $|l\rangle$ at time $\tau$ is calculated. Remembering Eqs. \eqref{smmm} and \eqref{smjj}, we find
\begin{eqnarray}
    {P_{l'l}(\tau)\over P_{ll'}(\tau)}=\Big[{1-J_3(\tau)\over J_2(\tau)}\Big]^{l-l'}=\Big[1+{1-|u(\tau)|^2\over v(\tau)}\Big]^{l-l'}.~
\end{eqnarray}
This interesting result implies that we can define an effective inverse temperature $\beta_\text{eff}(\tau)=\omega_0^{-1}\ln[1+{1-|u(\tau)|^2\over v(\tau)}]$, which is independent of $l$ and $l'$, such that the detailed balance
\begin{equation}
  {P_{l'l}(\tau)\over P_{ll'}(\tau)}=\exp[-\beta_\text{eff}(\tau)(E_{l'}-E_l)] \label{smeff}
\end{equation}is satisfied by our exact non-Markovian dynamics. This remarkable result goes beyond the weak-coupling condition required in the conventional detailed balance.

Substituting Eq. \eqref{smeff} into Eq.~(\ref{eq:eqs4}) and using $P_{l}(0)=e^{\beta_\text{s}\omega_0(l'-l)}P_{l'}(0)$, we have
\begin{eqnarray}
\chi_{\tau}(\xi)&=&\sum_{l',l}P_{ll'}(\tau)P_{l'}(0)e^{[\beta_{\text{eff}}(\tau)-\beta_{\text{s}}-\xi]\omega_{0}(l-l')}\nonumber\\
&=&\chi_{\tau}[\beta_{\text{eff}}(\tau)-\beta_{\text{s}}-\xi],\label{smsmt}
\end{eqnarray}
which is a direct manifestation of our generalized detailed balance condition in Eq. \eqref{smeff}. The combination of Eq. \eqref{smsmt} with the definition $\chi_\tau(\xi)=\int d\mathcal{Q} e^{\xi\mathcal{Q}}\mathcal{P}_\tau(\mathcal{Q})$ readily leads to ${\mathcal{P}_\tau(-\mathcal{Q})\over \mathcal{P}_\tau(\mathcal{Q})}=e^{[\beta_\text{eff}(\tau)-\beta_\text{s}]\mathcal{Q}}$. Integrating both sides over $\mathcal{Q}$ under the distribution $\mathcal{P}_\tau(\mathcal{Q})$ and using $\int d\mathcal{Q}\mathcal{P}_\tau(\mathcal{Q})=1$, we have
\begin{equation}
    \langle e^{[\beta_\text{eff}(\tau)-\beta_\text{s}]\mathcal{Q}}\rangle\equiv\int d\mathcal{Q}\mathcal{P}_\tau(\mathcal{Q})e^{[\beta_\text{eff}(\tau)-\beta_\text{s}]\mathcal{Q}}=1,
\end{equation}which is the generalized quantum fluctuation theorem for energy exchange in the non-Markovian dynamics.

\begin{figure}
\centering
\includegraphics[angle=0,width=0.475\textwidth]{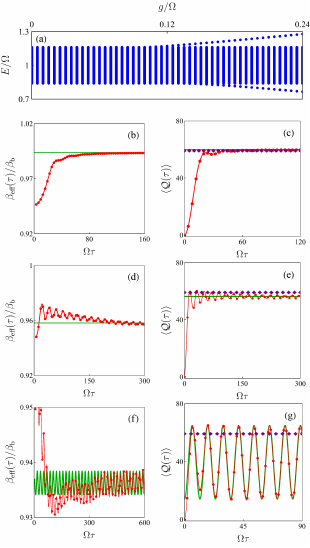}
\caption{(a) Energy spectrum of the total system. Evolution of $\beta_{\text{eff}}(\tau)/\beta_{\text{b}}$ and $\langle\mathcal{Q}(\tau)\rangle$ when $g/\Omega=0.08$ (b) and (c), $0.12$ (d) and (e), and $0.15$ (f) and (g). The red circles are exact numerical results. The green solid lines are analytical results from $u_{\text{TBS}}(\infty)$ and $v_{\text{TBS}}(\infty)$. The purple rhombuses are the Born-Markovian results in the long-time limit. We use $\beta_{\text{s}}=0.5\Omega^{-1}$, $\beta_{\text{b}}=0.2\Omega^{-1}$, $\omega_{0}=1.05\Omega$, and $\zeta=0.08\Omega$.}\label{fig:figsm}
\end{figure}

\section{Real-time evolution of the average energy exchange}
With the help of the reservoir engineering technique, the primary parameters in the spectral density are experimentally controllable. The spectral density of a quantum emitter coupled to the surface-plasmon polariton as a bath is adjustable by changing the distance between them~\cite{Andersen2011,PhysRevB.95.161408}. For a reservoir consisting of ultracold atomic gas, the Ohmicity parameter $s$ is tunable from the sub-Ohmic to the super-Ohmic forms by increasing the scattering length of the gas via Feshbach resonances~\cite{PhysRevA.84.031602}. These experimental progresses suggest that, via controlling the spectral density and the corresponding energy spectrum, we are able to realize a steerable nonequilibrium energy exchange in the platform of coupled-cavity arrays. As displayed in Fig.~\ref{fig:figsm}, the dynamics of the effective inverse temperature and the average energy exchange are sensitively depend on the structure of the energy spectrum. If the bound state is absent, then the effective inverse temperature and average energy exchange would tend to the corresponding values of the Born-Markovian approximate results. When one bound state is formed, they also tend to constant value, but shows a dramatic deviation from the Born-Markovian approximate ones. As long as two bound state are formed, both the effective inverse temperature and the average energy exchange show a lossless oscillation. This coherent behavior is a unique feature of the non-Markovian dynamics. Such a dynamically controllable energy exchange may have potential applications in the studies of finite-time quantum engines. In a finite-time quantum engine, the thermalization time is limited and the strokes of the working substance in contact with thermal reservoirs are commonly described by a non-Markovian master equation~\cite{HamedaniRaja_2021,Wiedmann_2020,PhysRevResearch.5.023066,PhysRevResearch.4.033233}. It is of importance to emphasize that our bound-state-based control scheme for energy exchange is completely different from that of Ref.~\cite{PhysRevX.10.031015}, which does not related to the characteristic of the energy spectrum.
	
\bibliography{reference}